\documentclass[submitted]{IEEEtran}
\usepackage{fancyhdr}
\usepackage{graphicx}
\usepackage{url}
\usepackage{cite}
\usepackage{color}
\usepackage{upgreek}
\usepackage[colorlinks=true,linkcolor=blue,urlcolor=blue,citecolor=blue]{hyperref}
\usepackage[colorinlistoftodos]{todonotes}

\begin{document}

\title{\bf Design of the Tsinghua Tabletop \\ Kibble Balance}
\author{Shisong Li$^{\dagger}$, \textit{Senior Member, IEEE}, Yongchao Ma, Wei Zhao,\\ Songling Huang, \textit{Senior Member, IEEE}, Xinjie Yu, \textit{Member, IEEE} 

\thanks{The authors are with the Department of Electrical Engineering, Tsinghua University, Beijing 100084, China.}
\thanks{$^\dagger$Email: shisongli@tsinghua.edu.cn}
\thanks{Submitted to \textit{IEEE Trans. Instrum. Meas.}}}

\maketitle

\begin{abstract}
The Kibble balance is a precision instrument for realizing the mass unit, the kilogram, in the new international system of units (SI). In recent years, an important trend for Kibble balance experiments is to go tabletop, in which the instrument's size is notably reduced while retaining a measurement accuracy of $10^{-8}$. In this paper, we report a new design of a tabletop Kibble balance to be built at Tsinghua University. The Tsinghua Kibble balance aims to deliver a compact instrument for robust mass calibrations from 10\,g to 1\,kg with a targeted measurement accuracy of 50\,$\upmu$g or less. Some major features of the Tsinghua Kibble balance system, including the design of a new magnet, one-mode measurement scheme, the spring-compensated magnet moving mechanism, and magnetic shielding considerations, are discussed. 

\end{abstract}


\section{Introduction}

\IEEEPARstart{T}{he} Kibble balance, originally known as the watt balance~\cite{Kibble1976}, is one of the two major approaches for realizing the unit of mass, the kilogram, in the new International System of Units (SI)~\cite{cgpm2018}. The other is known as the x-ray crystal density (XRCD) method~\cite{fujii2016realization}. At present, over a dozen groups, primarily national metrology institutes (NMIs), are conducting Kibble balance experiments~\cite{NRC,NIST,METAS,BIPM,LNE,MSL,NIM,KRISS,UME,PTB}, and the most accurate Kibble balances can calibrate masses at the kilogram level with a relative uncertainty of approximately one part in $10^8$~\cite{NRC,NIST}.


The goal of a Kibble balance is to establish a quasi-quantum mass standard by comparing mechanical power to electrical power \cite{haddad2016bridging}, and the latter can be precisely determined via quantum electrical standards, i.e., the Josephson voltage standard~\cite{tang201210} and the quantum Hall resistance standard~\cite{jeckelmann2001quantum}. 

A conventional Kibble balance operation contains two measurement phases: In the weighing phase, a current-carrying coil is placed in a magnetic field $B$, and the electromagnetic force is balanced by the weight of a test mass $m$, i.e. $mg=BlI$, where $g$ is the local gravitational acceleration, $l$ the coil wire length and $I$ the current through the coil. In the velocity phase, the geometrical factor $Bl$ is calibrated by moving the coil vertically with a velocity $v$ in the same magnetic field, and $Bl$ is given by the ratio of the induced voltage $U$ and the velocity $v$, i.e. $Bl=U/v$. By eliminating the same geometrical factor $Bl$ in two measurement phases, the mass is determined as $m=UI/(gv)$. A detailed principle of the Kibble balance experiment can be found in \cite{Stephan16}.

In general, highly accurate Kibble balances are relatively sizable experiments, oftentimes exceeding 1000\,kg \cite{NIST,NRC}. In recent years, several groups have started the development of tabletop-sized Kibble balances, e.g.~\cite{NPL,PTB,chao2020performance}. However, compact systems, especially those with small magnets, may encounter enlarged systematic effects \cite{li2022}. Therefore, it is a challenging task to minimize each related uncertainty and meanwhile realize a compact measurement apparatus. 

In this paper, we present the tabletop Kibble balance design at Tsinghua University. A core concern of the Tsinghua experiment is to find the balance between measurement accuracy and experiment compactness. Some ideas for this goal are discussed in detail following the summary presented in \cite{licpem2022}.

\section{General design goals}

\begin{figure}[tp!]
\centering
\includegraphics[width=0.5\textwidth]{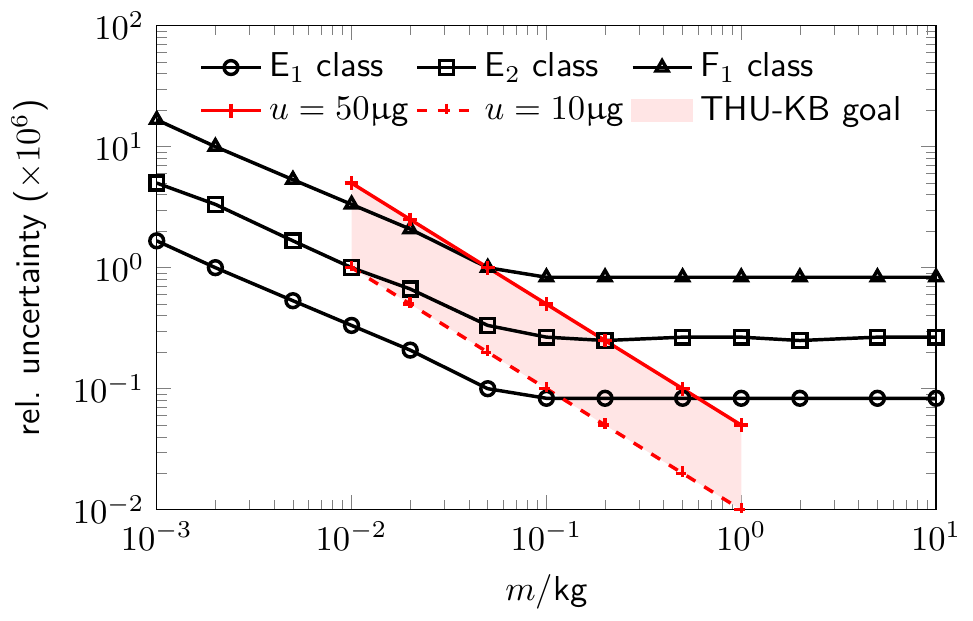}
\caption{Uncertainty for different mass standards and the measurement goal of the Tsinghua tabletop Kibble balance.}
\label{fig1}
\end{figure}

 Fig. \ref{fig1} shows the general goal of measurement accuracy for the Tsinghua tabletop Kibble balance. The three black lines are measurement uncertainties respectively for E$_1$, E$_2$ and F$_1$ class mass standards~\cite{oiml2004111}. The two red curves present the absolute measurement uncertainty of a Kibble balance with $10\,\upmu$g and $50\,\upmu$g over the range of 10\,g to 1\,kg. The Tsinghua tabletop Kibble balance aims to deliver a compact mass calibration machine with an accuracy of 50\,$\upmu$g or below in this range (10\,g to 1\,kg), and the measurement capacity is within the red shadow area. At the kilogram level, the instrument can calibrate the E$_1$ class mass standard, and at the 100\,g mass level, the measurement accuracy is comparable with the E$_2$ class mass standard. It is expected to calibrate F$_1$ class mass at the minimum measurement capacity, 10\,g. 

\begin{figure}[tp!]
\centering
\includegraphics[width=0.5\textwidth]{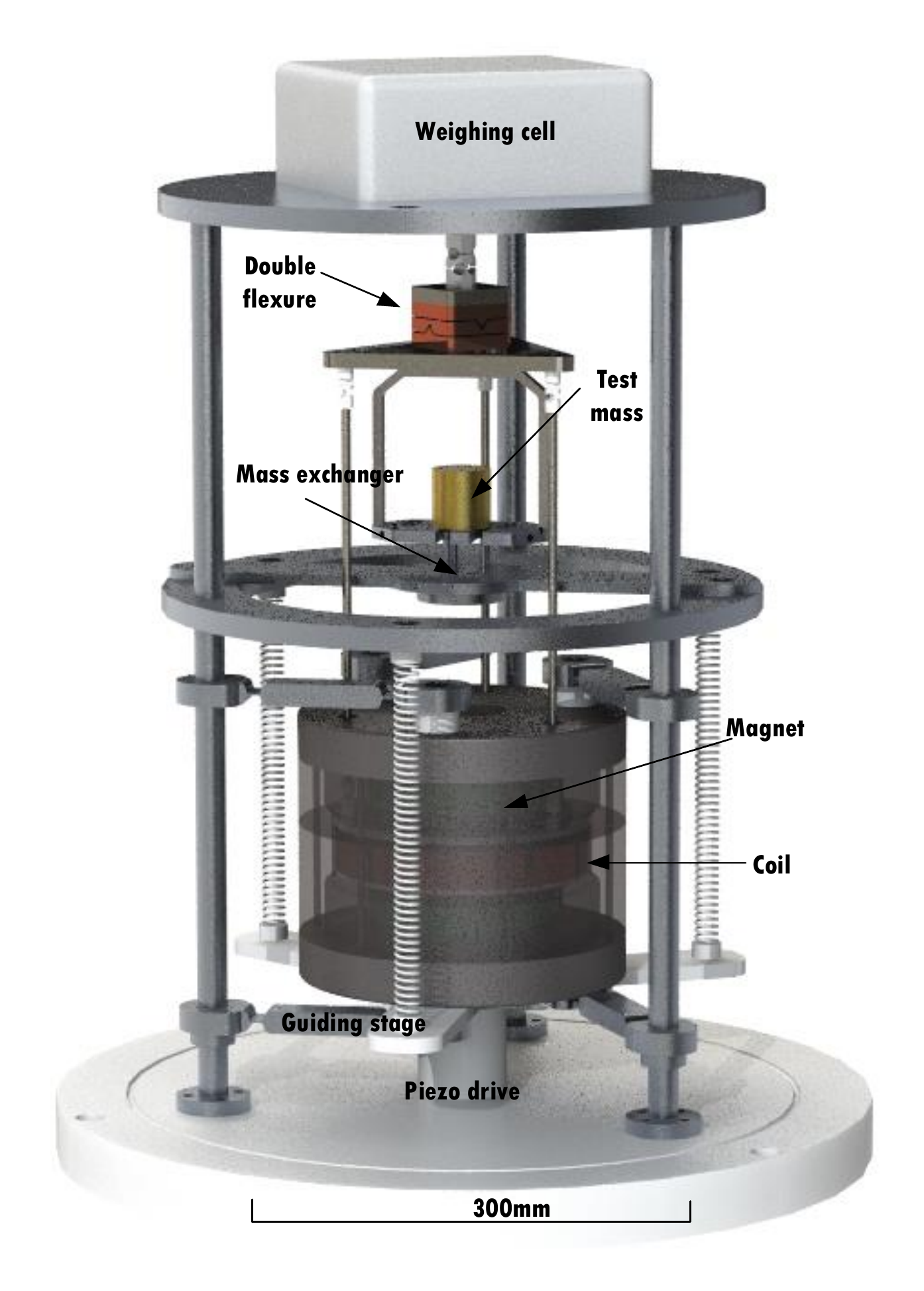}
\caption{Schematic drawing of the Tsinghua tabletop Kibble balance.}
\label{fig2}
\end{figure}

The overall design of the Tsinghua tabletop Kibble balance system is shown in Fig. \ref{fig2}. The outer enclosure, i.e. vacuum chamber (not shown in the plot), is designed approximately 0.5\,m in diameter and less than 1\,m in height. The total mass of the apparatus is expected to be within 100\,kg. Note that tabletop does not mean tiny. Based on the limitation between apparatus compactness and measurement accuracy~\cite{li2022}, the Tsinghua experiment will still put the measurement accuracy as a high-priority target, but not the only target. While within the relative uncertainty of a few parts in $10^8$, how to realize an apparatus as compact, robust, easy to operate, and cost-affordable as possible will be also a significant goal. Another important feature of the Tsinghua tabletop Kibble balance system is to go open hardware. By publishing the hardware design and software codes, we hope to offer modules for other experimenters. We are also interested to start an open Kibble balance platform through international collaborations, where experimenters in the field are encouraged to share, discuss, and exchange ideas or components for non-commercial purposes.

\section{Magnetic circuit design}
\begin{figure}[tp!]
\centering
\includegraphics[width=0.5\textwidth]{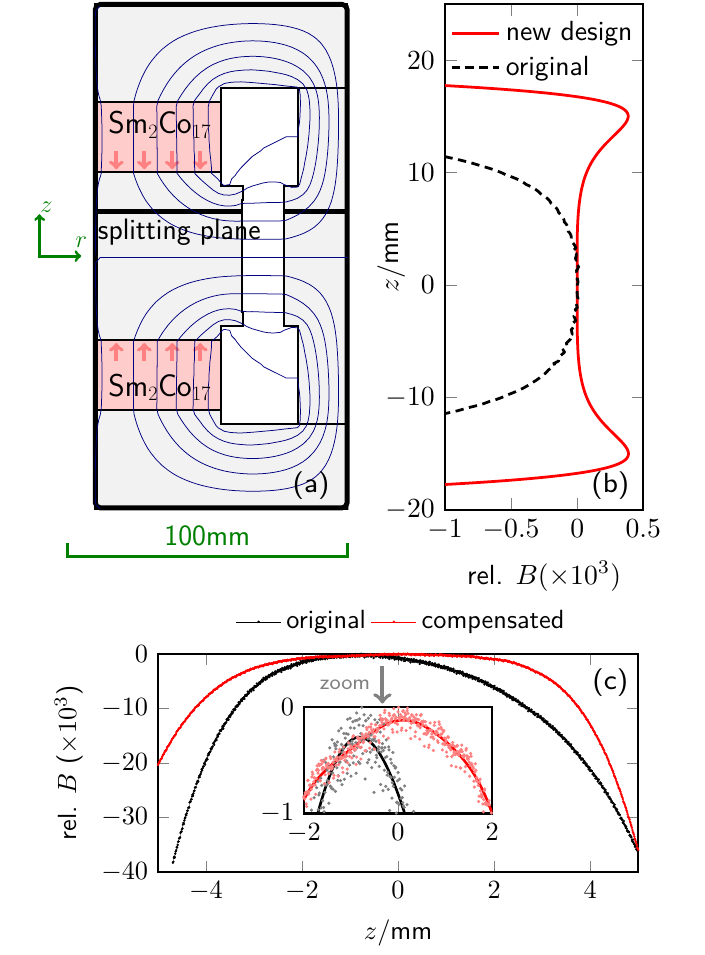}
\caption{(a) shows the magnetic circuit for Tsinghua tabletop Kibble balance system. The blue lines denote the magnetic flux. (b) compares the magnetic profiles of the original and modified BIPM-type magnetic circuits. {(c) presents an experimental comparison of the magnetic profiles of two demonstration magnet systems with/without compensation. The width and height of the air gap are 7.5\,mm and 15\,mm, and for the compensated magnet, an additional 0.2\,mm width, 2.5\,mm height rectangle is added in the inner yoke manufacture. The average magnetic flux density of two magnets is about 0.22\,T at the gap center. The profiles are measured by the gradient coil approach, see \cite{NISTmag}.}}
\label{fig3}
\end{figure}

Among different Kibble balance permanent magnet systems (see a detailed summary in \cite{li2022irony}), the Tsinghua experiment employs a BIPM-type magnetic circuit~\cite{NISTmag}. As shown in Fig.~\ref{fig3}(a), the yoke cylinder has an outer diameter of 220\,mm and a total height of 180\,mm. The flux of two 25\,mm-thick permanent magnets is radially distributed through an air gap with a cross-section of 15\,mm width and 50\,mm height. {The permanent magnet rings are made of Sm$_2$Co$_{17}$ with a typical temperature coefficient $\approx-3.6\times10^{-4}$/K.} The average magnetic flux density in the air gap is about 0.48\,T at the mean radius $r_c=80$\,mm. The total mass of the magnet is 38\,kg. 

Compared to conventional BIPM-type Kibble balance magnets, there are two noticeable features of the Tsinghua magnetic circuit design. The first is magnetic profile compensation. It is found in \cite{li2020simple} that the uniformity range of an air-gapped permanent magnet can be significantly improved by mechanically compensating two ends of the inner yoke with a minor additional yoke cylinder. This idea will be practically applied in the Tsinghua magnet manufacturing, and by Finite Element Analysis (FEA) an optimal shape of the compensation cylinder of width $=0.4$\,mm, height $=5$\,mm is found. Fig.~\ref{fig3}(b) compares the center magnetic profile $B(r_c,z)$ in the air gap before and after inner yoke modification. It can be seen that the modification almost doubles the flat magnetic profile range and hence can reduce the magnet height and volume. The design in this case can yield a usable flat profile for velocity measurement over 30\,mm. {To experimentally verify the compensation idea, two demonstration magnet systems with/without profile compensation, which are much smaller than the Tsinghua magnet system, are fabricated. The profiles, obtained by measuring the induced voltages on a gradient coil \cite{NISTmag}, are shown in Fig.\ref{fig3} (c). It can be seen that the proposed inner yoke compensation can significantly improve the uniformity of the magnetic profile, as predicted by the theoretical analysis.}

Schlamminger pointed out in \cite{schlamminger2013design} that the $Bl$ value of a Kibble balance should be neither too large nor too small, and the optimal is determined as
\begin{equation}
(Bl)_{op}=\sqrt{\frac{mgR}{v}},
\end{equation}
where $R$ is the sampling resistance for the current measurement in the weighing phase. The typical $Bl$ value, depending on the chosen $R$ and $v$, is usually about a few hundred Tm. To reach such a value, the wire gauge needs to be reduced in tabletop experiments, which however increases the coil ohmic loss in the weighing measurement and may bring considerable systematic effects~\cite{li2022}. A wider range of uniform magnetic profile allows a larger wire gauge, which reduces the coil resistance and hence is beneficial to suppress measurement bias related to coil heating. 

The second feature of the Tsinghua magnet system is the disassembly force reduction. 
The BIPM-type magnet has a merit of self-magnetic shielding, however, the drawback is that experimenters can not set the coil directly in the air gap. To insert or remove the coil, the yoke needs to be opened and if the splitting surface is not well chosen, the force (usually attractive) can easily reach a few kN~\cite{NISTmag}. Taking the above magnet circuit as an example, a splitting force of 6\,kN is required to open the magnet top/bottom cover, and a 3\,kN attraction force is to overcome if the splitting surface is set at the end of the air gap. 

\begin{figure}[tp!]
\centering
\includegraphics[width=0.5\textwidth]{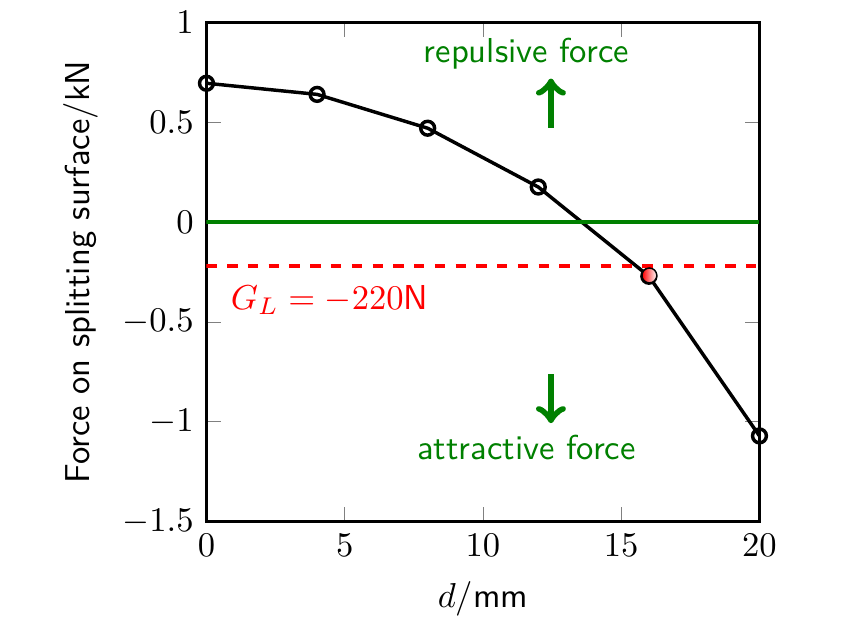}
\caption{The magnet open/close force as a function of the splitting surface position. $d$ is the distance between the splitting surface and the center of the magnet. }
\label{fig4}
\end{figure}

Fig.~\ref{fig4} presents the magnetic force required to open/close the magnetic circuit as a function of $d$, the distance of the splitting surface to the center of the magnet. With $d=0$\,mm, the splitting force is about 700\,N repulsive. This can be explained by conjugating two identical LNE-magnet systems~\cite{LNEmag}: the magnetic flux bends at each air gap end and since they have the same magnetic pole, the magnetic force is hence repulsive. The reversal of the splitting force direction means that an optimal with zero splitting force exists. In Fig.~\ref{fig4}, when $d=13.8$\,mm, the splitting force is minimum. However, from Fig.~\ref{fig3}, $d=13.8$\,mm is still in the uniform field range. In order to lower the magnetic profile change due to the splitting plane and gain as wide a uniform field range as possible, the Tsinghua magnet is designed to be split at a surface 16\,mm higher from the symmetrical plane ($z=0$). The FEA simulation shows the disassembly force at this position is -270\,N (attractive). When the magnetic circuit is required to open, the top part of the magnet (16\,kg) will be fixed, and the attractive force is compensated by the weight of the lower part ($\approx 220$\,N), resulting in a net force of about 50\,N. The residual attractive force ensures the lower magnet part can stay even without a connecting screw. By taking advantage of the magnet gravity, the magnetic profile is also less affected. Since the residual open/close force is low, this design also brings added benefits such as directly installing or removing the coil without needing to fully extract the magnet.   

\section{One-mode measurement scheme}
When the magnet volume is reduced, systematic effects such as the current effect \cite{li17}, the nonlinear magnetic error \cite{hysteresis}, and the difference in thermal dissipation between weighing and velocity measurements \cite{shisong2022} may introduce considerable uncertainty components. These effects scale mostly quadratically with the current in the coil, i.e., $\beta I^2$, and hence, cannot be eliminated by mass-on and mass-off measurements. To reduce these terms to a negligible level, the one-mode, two-phase measurement (OMTP) scheme similar to the BIPM balance operation~\cite{BIPM} will be used in the Tsinghua tabletop Kibble balance. {As a reminder, it was suggested by \cite{Stephan16} and clarified in \cite{li17} that the word "mode" refers to the number of current on/off states in the measurement: In the two-mode scheme, the current is on during the weighing measurement and off during the velocity measurement, while in the one-mode scheme, the current is on during both the weighing and velocity measurements. The word "phase" refers to the timing of the measurement: two-phase means that the weighing and velocity measurements are separated in time, while in one-phase the weighing and velocity measurements are simultaneous.}

\begin{figure}[tp!]
\centering
\includegraphics[width=0.45\textwidth]{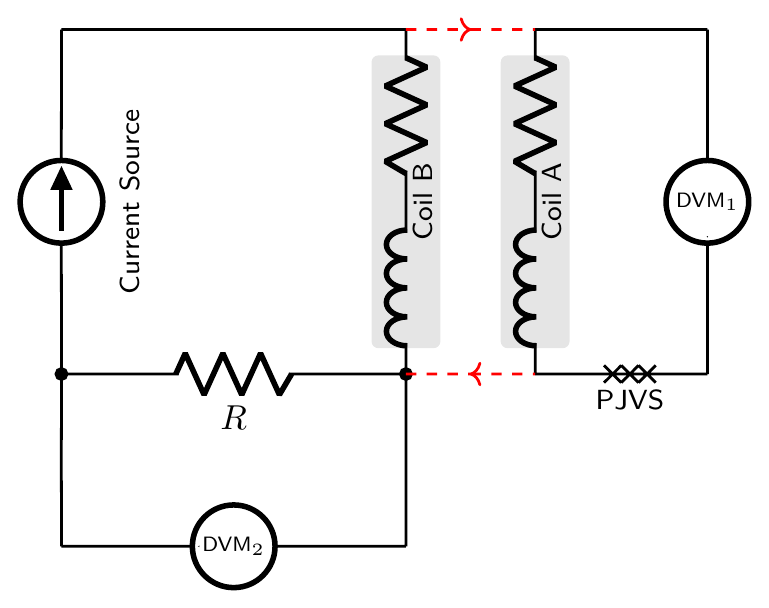}\\
(a) velocity phase\\
\includegraphics[width=0.45\textwidth]{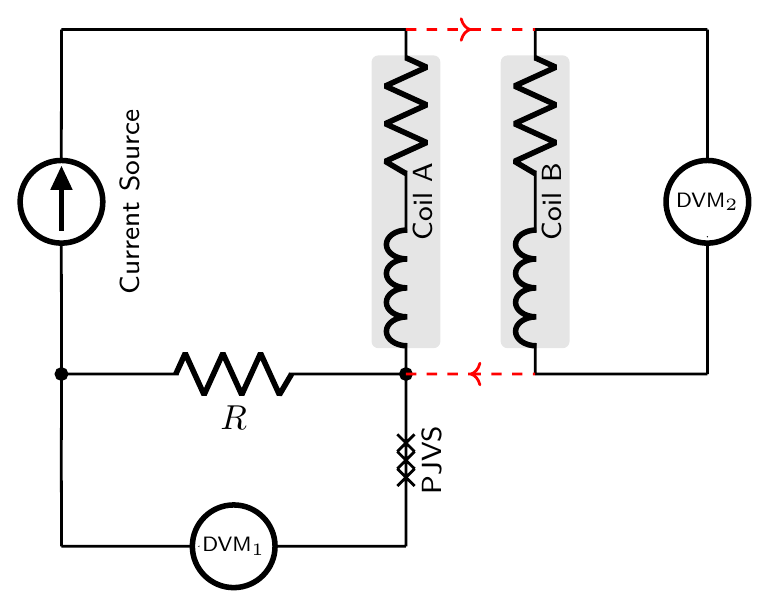}\\
(b) weighing phase
\caption{Electrical circuit for one-mode, two-phase measurement scheme. (a) and (b) are respectively for the velocity and weighing measurements. $R$ is the sampling resistance for the current measurement. PJVS denotes the programmable Josephson voltage standard. Coil A and coil B are two windings of the bifilar coil. }
\label{fig5}
\end{figure}

Fig.~\ref{fig5} presents a typical electrical circuit for the OMTP measurement scheme. Compared to the conventional two-mode, two-phase (TMTP) measurement scheme, the OMTP scheme employs a bifilar coil (coil A and coil B) so that the current in two measurement phases can stay symmetrical. In the velocity phase, coil B
carries the current and coil A for the $U/v$ measurement. In the weighing phase, coil A will hold
the current to ensure the $Bl$ measured is from the same coil. Different from the TMTP scheme, the OMTP contains two measurement steps in the velocity phase, with respectively plus current and minus current in coil B. This arrangement follows the two-step measurement in the
weighing phase, i.e. mass-on and mass-off, and keeps the magnetic status as close as possible for the two measurement phases.

\begin{figure}[tp!]
\centering
\includegraphics[width=0.5\textwidth]{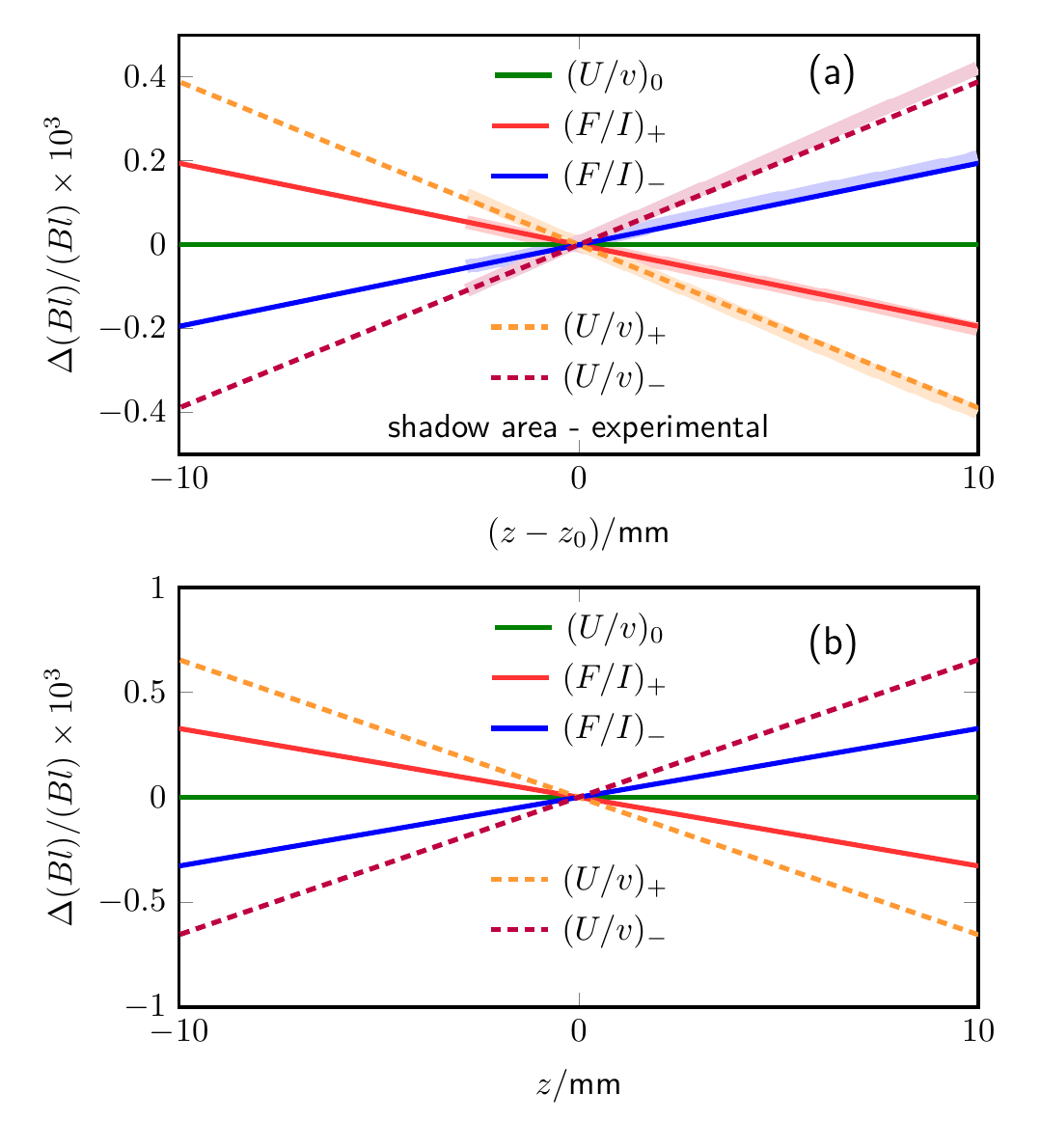}
\caption{Magnetic profile change related to the coil inductance effect. (a) shows the profile comparison of experimental data and FEA result in the BIPM system~\cite{li17}. The subscripts, $+, -$ and 0 denote respectively the plus, minus and zero current profiles. The thin lines are based on FEA calculation while the shadow area is the experimental measurement. (b) presents the estimated $Bl$ change of the Tsinghua tabletop system. The total ampere-turns through the coil is 20.22\,A for generating a 4.9\,N magnetic force. The weighting profile change is obtained by the FEA calculation of the inductance force. }
\label{figx}
\end{figure}

Although the OMTP measurement scheme can well suppress the bias related to nonlinear and thermal magnetic effects, the following major concerns should be considered and addressed. The first is the coil inductance effect, i.e. an additional magnetic profile term generated by the coil flux during the velocity measurement must be considered. It is shown in \cite{li17,li18} that the
component added to the magnetic profile is approximately a linear function of the coil vertical position $z$ in an air-gapped magnet system, caused by the coil inductance energy change. { For the weighing, the $Bl$ change can be directly deduced by the virtual principle, i.e.
\begin{equation}
    \Delta(Bl)_w=\frac{I}{2}\frac{\partial L}{\partial z}.
\end{equation}
where $L$ denotes the inductance of the current-carrying coil.}
While with consideration of the coil inductance, the total induction of the emf-measurement coil is written as
\begin{equation}
U(I)=Blv+\frac{\partial (MI)}{\partial t},
\label{eq2}
\end{equation}
where $M$ is the mutual inductance between two coils. Note that for a bifilar coil, the inductance of each coil $L$ approximately equals their
mutual inductance, i.e. $L=M$, because they share the same coil flux at a certain vertical position. Using $v=\partial z/\partial t$, (\ref{eq2}) can be rewritten as
\begin{equation}
\Delta(Bl)_v=\frac{U(I)}{v}-Bl=I\frac{\partial L}{\partial z}+L\frac{\partial I}{\partial z}.
\label{eq3}
\end{equation}
In (\ref{eq3}), two additional terms, $I({\partial L}/{\partial z})$ and $L({\partial I}/{\partial z})$, are introduced to the magnetic profile measurement in the velocity phase. The second additional term $L({\partial I}/{\partial z})$ in principle can be removed by using a constant current in the velocity measurement. However, the current variation during velocity measurement introduces electrical noise for the induced voltage measurement, and hence a quiet current source is required. {As shown in Fig. \ref{figx} (a), how the $\partial L/\partial z$ term influences the weighing and velocity measurements has been well understood,  both experimentally and numerically \cite{li17}. The general conclusion is that with the same current, the velocity profile change related to the coil self/mutual inductance is twice that of the weighing measurement, i.e. $\partial (\Delta Bl)_v/\partial z=2\partial (\Delta Bl)_w/\partial z$. Fig. \ref{figx} (b) presents the relative profile changes related to coil inductance for the Tsinghua tabletop system. The slopes are about 50\% higher than the BIPM system. }

The parameters of the bifilar coil employed in the Tsinghua tabletop Kibble balance are designed as follows: The diameter of the copper wire is chosen 0.2\,mm. In a 10\,mm$\times$10\,mm section area, the total number of wire turns $N=({10}/{0.2})^2=2500$ is obtained. For each winding, the wire turns should be $N/2=1250$. Taking $r_c=80$\,mm, $B=0.48$\,T, $Bl=2\pi r_c (N/2) B\approx300$\,Tm. In this case, the coil resistance $R_c$ is approximately 340\,$\Omega$. Note that the above estimation is based on a 20\,mm uniform magnetic field range, in which 10\,mm is used for the coil movement trajectory. The $Bl$ value can be adjusted depending on the actual profile flatness, and in the ideal case (30\,mm usable range) shown in Fig.~\ref{fig3}(b), the $Bl$ can reach a maximum value of 600\,Tm for each single winding.  

Another concern for the OMTP scheme is an electrical leakage issue for the bifilar coil. The electrical leakage of a TMTP Kibble balance can be directly measured and controlled to a level of a few parts in $10^9$ or lower. In OMTP scheme, the leakage between two windings of the bifilar coil is unique and should be considered. Reference \cite{BIPMmag2017} reported a measurement of the leakage resistance $R_L$ between two windings in a bifilar coil, showing $R_L$ is only a few \,G$\Omega$. In this case, a leakage current $i$, the dotted red lines presented in Fig.~\ref{fig5}, will go through the none-current-carrying coil. In the velocity measurement, the additional emf drop related to the leakage is similar to the thermal voltage, and can be removed by averaging the upward and downward measurements. The relative amplitude of this term can be estimated as $R_c^2I/(R_LU)$, and in the above example it is a few parts in $10^6$, comparable to the thermal voltage value. In the weighing measurement, the leakage current from coil A to coil B will also generate magnetic force, and hence the weighing effect is minor. 

As an additional advantage, the one-mode measurement scheme simplifies the mass exchange as the electromagnetic force is strong enough to be used as a holding force. Furthermore, no tare mass is required in the velocity measurement.

\section{Magnet-moving mechanism}
Kibble and Robinson showed that using the same mechanism for weighing and velocity measurements can relax the demands on alignment~\cite{kibble2014principles}. The beam balance~\cite{NRC} and wheel balance~\cite{NIST} accomplish that, but, both require a large volume to realize good linearity and sensitivity. The Tsinghua system will use, similar to other Kibble balances, a weighing cell. The main difference, however, is that here the magnet is moved and not the coil or even the weighing unit. Without moving the sensitive parts of the system, a more precise weighing can be made. The disadvantages of moving the magnet and solutions in the Tsinghua experiment are discussed as follows.

\begin{figure}[tp!]
\centering
\includegraphics[width=0.5\textwidth]{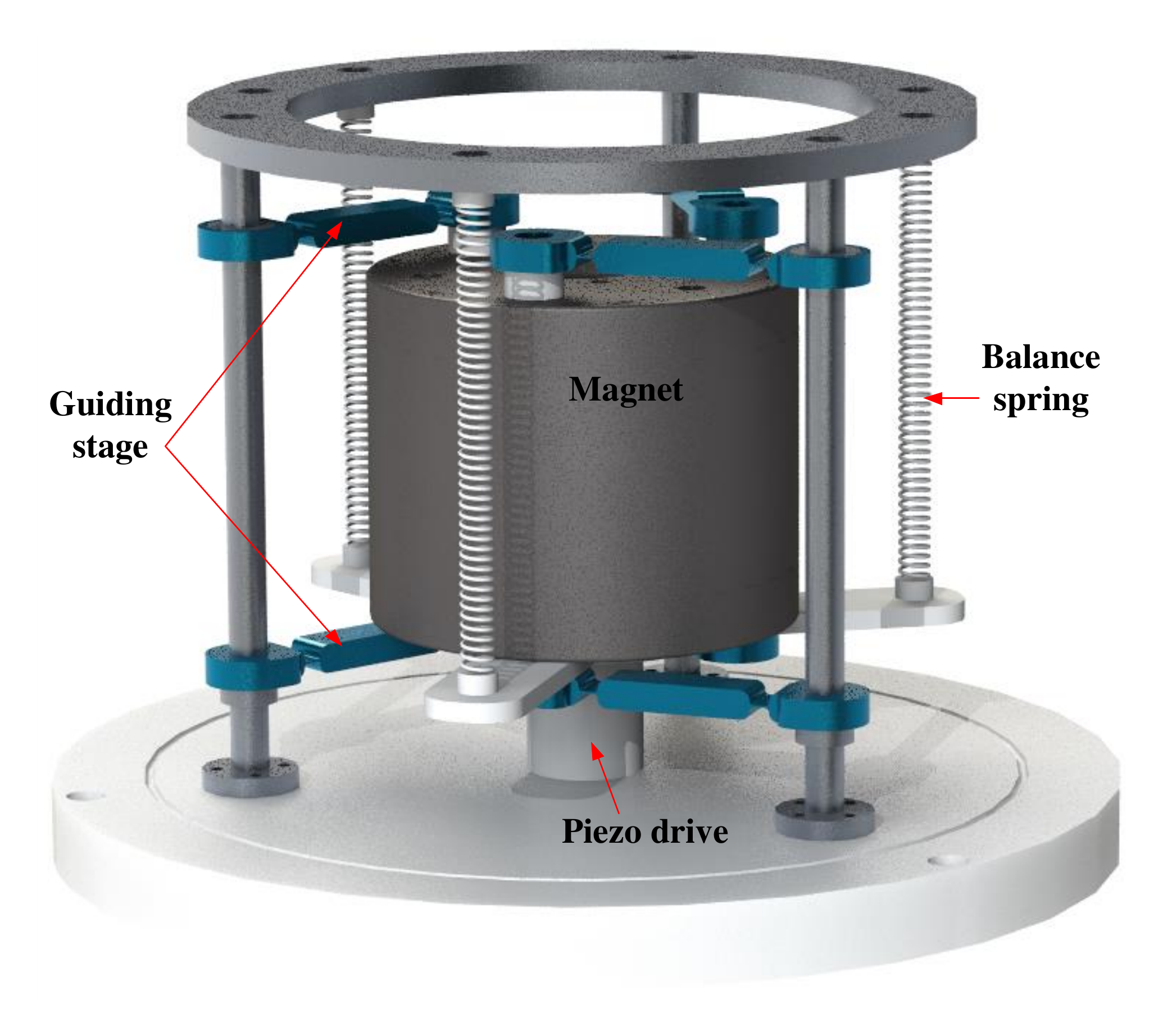}
\caption{A schematic drawing of the magnet-moving stage.}
\label{fig6}
\end{figure}

For a magnet-moving system, a very strong motor is required to move the heavy magnet. To reduce the driving force required, as shown in Fig.\ref{fig6}, the Tsinghua Kibble balance design employs a seismometer-style linear guidance mechanism to move the magnet. Three springs are used to balance the weight of the magnet $G$, yielding $G=3k\Delta z$ where $k$ is the spring constant and $\Delta z$ the spring length change. Three or six pairs of flexure bars are used to guide the movement to ensure almost perfect vertical motion. The design sets $\Delta z$ to 100\,mm, and the magnet moving range is targeted to be $\pm5$\,mm. With these parameters, a total driving force of $\pm G/20\approx\pm20$\,N is required.  

\begin{figure}[tp!]
\centering
\includegraphics[width=0.5\textwidth]{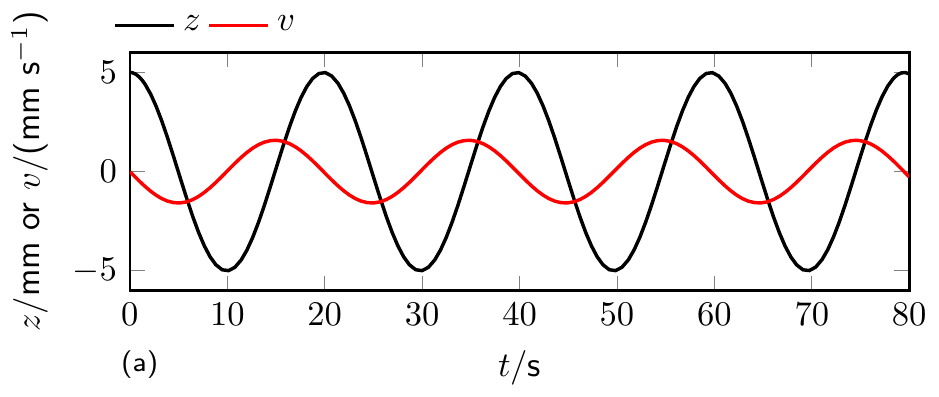}\\
\includegraphics[width=0.5\textwidth]{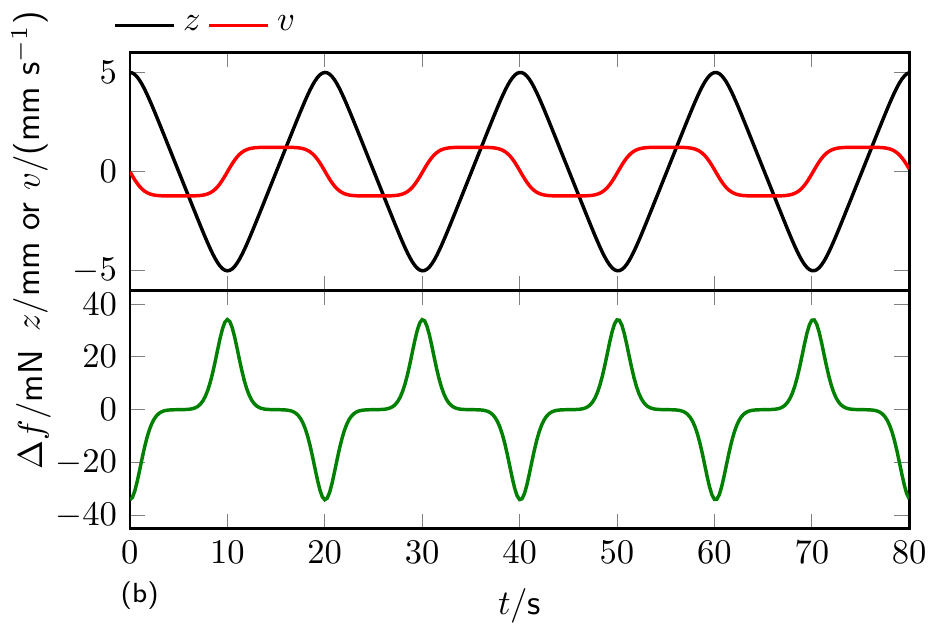}
\caption{(a) presents a sample oscillator with oscillation period $T\approx20$\,s. (b) is a nonlinear oscillator with a uniform velocity range. The lower plot is the residual force with the spring restore force fully compensated.}
\label{fig7}
\end{figure}

For magnet-moving, different driving strategies are available. The easiest way is to drive the stage linearly. Linear stages with a load capacity of a few kilograms, such as piezo motors, can be used to drive the movement of the magnet. The second method is using periodic oscillators. The spring system is a natural harmonic oscillator written as
\begin{equation}
\mathcal{M}\frac{\mathrm{d}^2z}{\mathrm{d}t^2}
+(3k-k_d)z=0,
\label{eq4}
\end{equation}
where $\mathcal{M}$ denotes the mass of the magnet, $k$ the spring constant, and $f_d=k_dz$ the drive force supplied by the motor. 
The static equation of the above spring system is written as $\mathcal{M}g=3k\Delta z$. Then, (\ref{eq4}) can be normalized as
\begin{equation}
\frac{\mathrm{d}^2z}{\mathrm{d}t^2}+\left(\frac{g}{\Delta z}-\frac{k_d}{\mathcal{M}}\right)z=0.
\label{eq5}
\end{equation}
First, let $k_d=0$. Substituting $\Delta z=100$\,mm, $g\approx 9.8$m/s$^2$ into (\ref{eq5}), it yields the oscillation period $T=2\pi\sqrt{\Delta z/{g}}\approx 0.63$\,s. The natural oscillation of the magnet-suspension system, in this case, generates a maximum moving speed of 50\,mm/s with a trajectory length of 10\,mm. Obviously, the oscillation speed is too fast for the velocity phase measurement. To lower the magnet moving speed to a typical value, e.g. a few mm/s, the driving force must supply compensation for the spring restore forces. Fig.~\ref{fig7}(a) presents an example with $T\approx20$\,s and the maximum $v=1.5$\,mm/s with a linear drive. In this case, the total restore term $g/\Delta z-k_d/\mathcal{M}\approx$0.1\,s$^{-2}$ is about two magnitudes smaller than $g/\Delta z$, and the driving force required can be considered approximately equal to the spring restore force, i.e. $\pm$20\,N in $\pm5$\,mm travel range. Here we also introduce a nonlinear oscillator to achieve an easier $U/v$ measurement. The idea is to make the velocity more constant during the movement by introducing nonlinear driven force terms, such as $z^3$, $z^5$ and higher odd terms. Fig.~\ref{fig7}(b) shows an example: the linear term in (\ref{eq5}) is replaced by $k_1z+k_3z^3+k_5z^5$ where $k_1=0.0012$\,s$^{-2}$, $k_2=615$\,s$^{-2}$m$^{-2}$, $k_3=2.46\times10^{8}$\,s$^{-2}$m$^{-4}$, -5\,mm$\leq z\leq$5\,mm. It can be seen from Fig.~\ref{fig7}(b) that in this case the velocity uniformity during a single movement is significantly improved, and experimenters can use only the constant velocity trajectory for $U/v$ measurement. The DC voltage measurement in such a scheme is somewhat easier and more accurate than AC voltage determination in conventional harmonic oscillators. The residual force, $\Delta f=(k_1z+k_3z^3+k_5z^5)\mathcal{M}$, is at the $10^{-2}$\,N level. {If the weight of the magnet is well compensated, generating an additional driving force component following $\Delta f(z)$ can realize the proposed nonlinear oscillation. Since $\Delta f$ is at the gram level, a compact actuator similar to the main coil-magnet system can be used, and the force generation, in this case, can be achieved by tracking the magnet position $z$ and digitally converting it into the required driving current.  }

Another drawback of the magnet-moving system is that the measurement is susceptible to the background magnetic field \cite{UME2,nimexmag}. In the weighing measurement, the $Bl$ is written as
\begin{equation}
\frac{mg}{I}=Bl+B_0l,
\label{exw}
\end{equation}
where $B_0$ is the coupled static field at the coil position from the background magnetic flux. 
While the velocity measurement is written as
\begin{equation}
\frac{U}{v}=Bl+B_0'l.
\label{exv}
\end{equation}
{where $B_0'$ is the coupled magnetic flux density in the velocity measurement. It can be seen by comparing (\ref{exw}) and (\ref{exv}) that the measurement error depends on $B_0'-B_0$. During the velocity measurement, the background magnetic flux and hence the coupled field at the coil position $B_0'$ redistributes when the magnet is moved to different vertical positions, $z$. If the background flux sources are static and identical in two measurement phases, it yields $B_0'(z_w)=B_0(z_w)$ where $z_w$ is the weighing position, and the coupled magnetic field component can be well canceled. However, in reality, the magnet moving mechanism introduces new flux sources and $B_0'$ and $B_0$ are no longer equal at the weighing position. Then it results in an additional contribution to the Faraday voltage. To ensure the measurement accuracy, the relative background flux coupling at the coil position, i.e. $(B_0'-B_0)/B$, should be lower than $1\times10^{-8}$, and hence good magnetic shielding is required.}

\begin{figure}[tp!]
\centering
\includegraphics[width=0.5\textwidth]{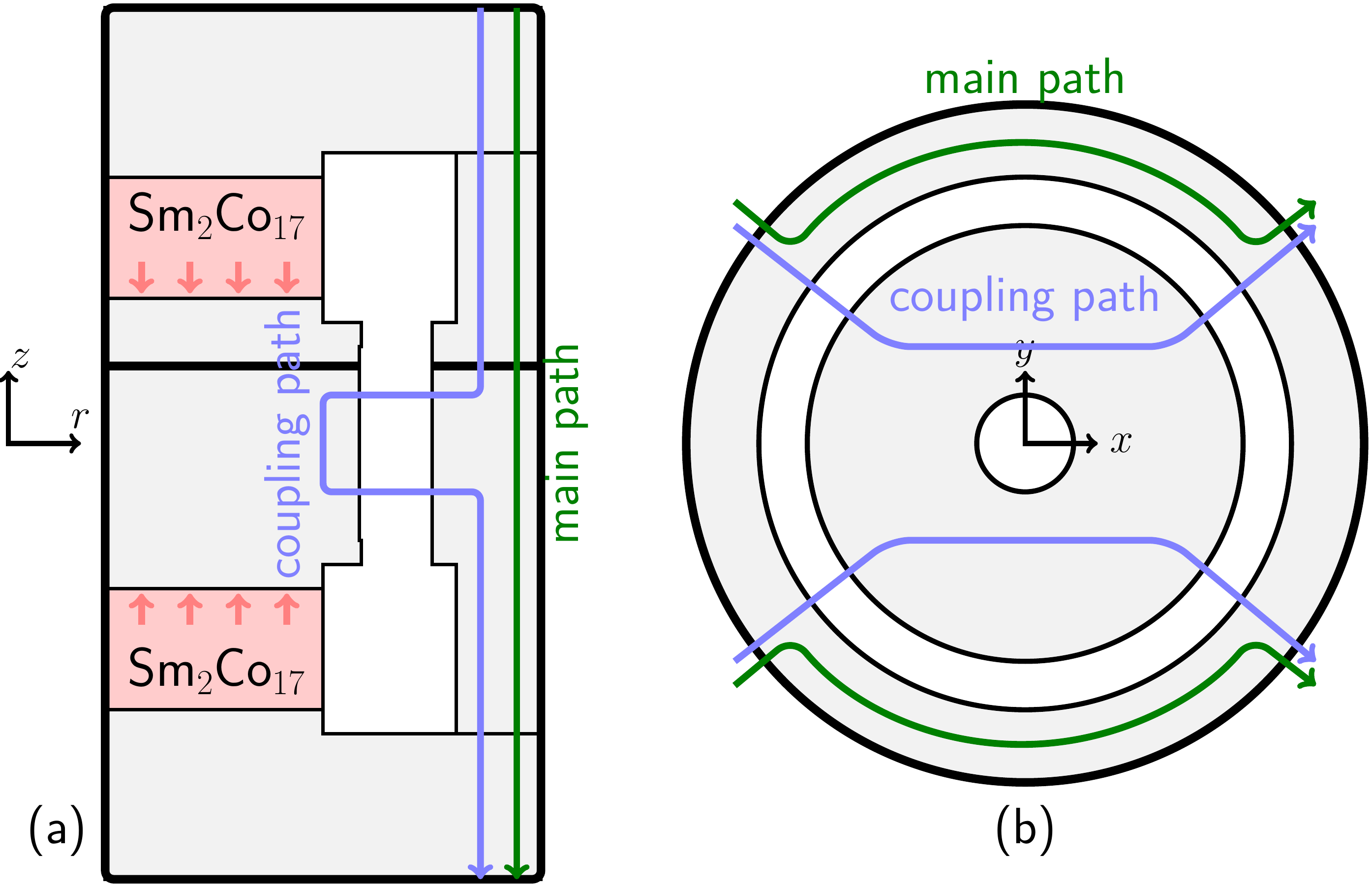}
\caption{Different magnetic flux coupling paths through the BIPM-type Kibble balance magnet. The background flux is (a) vertical and (b) horizontal.}
\label{fig8}
\end{figure}

Several measures will be used in the Tsinghua tabletop system to reduce the bias caused by external magnetic flux. As a novel design aspect, we plan to use low-carbon steel as a material for the vacuum chamber. The chamber will double as a magnetic shield, keeping external magnetic fields, {especially time-varying magnetic flux sources,} away from the coil. Together with the good self-shielding performance of the BIPM-type magnet \cite{BIPMmagShielding}, the far-field interference can be well rejected. It is worth noting how the BIPM-type magnet keeps the far field away from the air gap. As shown in Fig.~\ref{fig8}(a), the majority of the vertical external flux will go through the outer yoke, and only a small amount of flux goes through the air gap (twice). FEA simulations show that the coupling magnet field $B_0$ is a linear function of $z$. Therefore, the coupling magnetic field can be significantly reduced by setting the weighing position at the magnet center, i.e. $z=0$. For the horizontal background flux, the main flux will pass through the outer yoke and top/bottom covers, the coupling field as shown in Fig.~\ref{fig8}(b) slightly strengthens the air gap magnetic field on one side while weakening the other side. This change will be compensated during the alignment and will not bring additional systematic effects. Note that the above analysis works only for far fields and it is not true for close-by magnetic sources. {Additional magnetic sources in the velocity measurement can be avoided by using for example piezo or electrostatic motors. If magnetic actuators are required, additional shields must be installed.}

{ 

When installing magnetic shielding, it is crucial to evaluate the residual static magnetic field coupled to the coil position. Although the experimental process can be challenging, it is not impossible. For example, the first step is to pre-calibrate the external magnetic field attenuation through the self-shielding of magnetic circuit. This can be achieved by introducing a known external flux source $B_E$ and calibrating the magnetic field change $\Delta B_a$ at the coil position as a function of $B_E$, yeilding $\gamma(B_E)=B_E/\Delta B_a$. The gain $\gamma$ at a smaller range can be extrapolated based on the measurement. Next, the magnetic field inside the vacuum chamber $B_e$ can be monitored using magnetic sensors such as a fluxgate with mT range and $\upmu$T sensitivity. By mapping the field in space, the residual coupled magnetic field at the coil position can be estimated as $B_e/\gamma(B_e)$. This estimation can be used to determine the accuracy of the shielding and to identify any necessary adjustments to further improve the performance of the system.

}

\section{Summary}
This paper introduces the main design ideas of the tabletop Kibble balance system at Tsinghua University. A compact magnet system is proposed to supply a long-range uniform magnetic field. Optimization of the splitting plane makes it easy to open and close the magnet. A one-mode, two-phase scheme is used to minimize measurement biases from currents and thermal loading. Related issues, such as the current effect, current noise problem, and bifilar coil leakage, are discussed. The conventional coil movement is replaced by a stage that moves the magnet assisted by springs. A few magnet-driving mechanisms are presented. The BIPM-type magnet circuit has good self-magnetic shielding and the vacuum enclosure doubles as a magnetic shield. The proposed design will reach measurement uncertainties at or below 50$\upmu$g for mass calibrations ranging from 10\,g to 1\,kg.

\section*{Acknowledgement}
The authors would like to thank Stephan Schlamminger, Leon Chao, Darine Haddad, David Newell, Frank Seifert from NIST, and Zhengkun Li, Yang Bai, Ping Yang from NIM for valuable discussions on the open-hardware idea. This work has been supported by the national key research and development program (2022YFF0708600).


\begin{thebibliography}{10}

\bibitem{Kibble1976}
B.~P. Kibble, ``A measurement of the gyromagnetic ratio of the proton by the
  strong field method,'' in {\em Atomic masses and fundamental constants 5},
  pp.~545--551, Springer, 1976.

\bibitem{cgpm2018}
 Resolution 1 of the 26th CGPM Conference, 2018.

\bibitem{fujii2016realization}
K.~Fujii, H.~Bettin, P.~Becker, {\it et al}, ``Realization of the kilogram by the XRCD method,'' {\em Metrologia}, vol.~53, no.~5, pp.~A19--A45, 2016.

\bibitem{NRC}
B.~M. Wood, C.~A. Sanchez, R.~G. Green, and J.~O. Liard, ``A summary of the
  Planck constant determinations using the NRC Kibble
  balance,'' {\em Metrologia}, vol.~54, no.~3, pp.~399--409, 2017.

\bibitem{NIST}
D.~Haddad, F.~Seifert, L.~S. Chao, {\it et al}, ``Measurement of the Planck constant at the National Institute of Standards and Technology from 2015 to 2017,'' {\em Metrologia}, vol.~54, no.~5, pp.~633--641, 2017.

\bibitem{METAS}
H.~Baumann, A.~Eichenberger, F.~Cosandier,  {\it et al}, ``Design of the new METAS watt balance experiment Mark II,'' {\em Metrologia}, vol.~50, no.~3, pp.~235--242, 2013.

\bibitem{BIPM}
H.~Fang, F.~Bielsa, S.~Li, A.~Kiss, and M.~Stock, ``The BIPM Kibble balance for realizing the kilogram definition,'' {\em Metrologia}, vol.~57, p.~045009, 2020.

\bibitem{LNE}
M.~Thomas, D.~Ziane, P.~Pinot,  {\it et al}, ``A determination of the Planck constant using the LNE Kibble balance in air,'' {\em
  Metrologia}, vol.~54, no.~4, pp.~468--480, 2017.

\bibitem{MSL}
C.~M. Sutton and M.~T. Clarkson, ``A magnet system for the MSL watt
  balance,'' {\em Metrologia}, vol.~51, no.~2, pp.~S101--S106, 2014.

\bibitem{NIM}
Z.~Li, Z.~Zhang, Y.~Lu, {\em et~al.}, ``The first determination of the Planck constant with the joule balance {NIM}-2,'' {\em Metrologia}, vol.~54, no.~5, pp.~763--774, 2017.

\bibitem{KRISS}
D.~Kim, B.-C. Woo, K.-C. Lee, {\em et al},
  ``Design of the {KRISS} watt balance,'' {\em Metrologia}, vol.~51,
  no.~2, pp.~S96--S100, 2014.

\bibitem{UME}
H.~Ahmedov, N.~B. A{\c{s}}k{\i}n, B.~Korutlu, and R.~Orhan, ``Preliminary
  {P}lanck constant measurements via {UME} oscillating magnet
  kibble balance,'' {\em Metrologia}, vol.~55, no.~3, pp.~326--333, 2018.

\bibitem{PTB}
C.~Rothleitner, J.~Schleichert, N.~Rogge, {\em et al}, ``The
  Planck balance using a fixed value of the Planck constant to calibrate
  E1/E2-weights,'' {\em Meas. Sci. Technol.}, vol.~29, no.~7,
  p.~074003, 2018.

\bibitem{haddad2016bridging}
D.~Haddad, F.~Seifert, L.~S. Chao, {\em et al}, ``Bridging classical and quantum mechanics,'' {\em Metrologia}, vol.~53, no.~5, pp.~A83--A85, 2016.

\bibitem{tang201210}
Y. Tang, V.~Ojha, S.~Schlamminger, {\em et al}, ``A 10V programmable Josephson voltage standard and its applications for voltage metrology,'' {\em Metrologia}, vol.~49,
  no.~6, p.~635, 2012.

\bibitem{jeckelmann2001quantum}
B.~Jeckelmann and B.~Jeanneret, ``The quantum {H}all effect as an
  electrical resistance standard,'' {\em Rep. Prog. Phys.}, vol.~64, no.~12,
  p.~1603, 2001.

\bibitem{Stephan16}
I.~A. Robinson and S.~Schlamminger, ``The watt or {K}ibble balance: a
  technique for implementing the new {SI} definition of the unit of
  mass,'' {\em Metrologia}, vol.~53, no.~5, pp.~A46--A74, 2016.

\bibitem{NPL}
I.~A. Robinson, ``Towards the redefinition of the kilogram: a measurement of
  the {P}lanck constant using the {NPL} {M}ark {II}
  watt balance,'' {\em Metrologia}, vol.~49, no.~1, pp.~113--156, 2011.

\bibitem{chao2020performance}
L.~Chao, F.~Seifert, D.~Haddad, {\em et al}, ``The performance of the KIBB-g1 tabletop Kibble balance at NIST,'' {\em
  Metrologia}, vol.~57, no.~3, p.~035014, 2020.

\bibitem{li2022}
S.~Li and S.~Schlamminger, ``Magnetic uncertainties for compact kibble
  balances: An investigation,'' {\em IEEE Trans. Instrum. Meas.}, vol. 71, pp. 1-9, 2022, Art no. 1502409.

\bibitem{licpem2022}
S.~Li, W.~Zhao,S. Huang, X. Yu ``Design of the Tsinghua open-hardware tabletop Kibble balance,'' in {\em 2022 Conference on Precision Electromagnetic Measurements (CPEM 2022)}, pp.~1--2, IEEE, 2022.
  
\bibitem{oiml2004111}
R.~OIML, ``111-1, weights of classes E1, E2, F1, F2, M1, M1-2, M2, M2-3 and M3 part 1: Metrological and technical requirements,'' {\em International
  Organization of Legal Metrology}, 2004.

\bibitem{li2022irony}
S.~Li and S.~Schlamminger, ``The irony of the magnet system for Kibble
  balances—a review,'' {\em Metrologia}, vol.~59, no.~2, p.~022001, 2022.

\bibitem{NISTmag}
F.~Seifert, A.~Panna, S.~Li, {\em et al}, ``Construction, measurement, shimming, and
  performance of the {NIST-4} magnet system,'' {\em IEEE Trans. Instrum.
  Meas.}, vol.~63, no.~12, pp.~3027--3038, 2014.

\bibitem{li2020simple}
S.~Li, S.~Schlamminger, and Q.~Wang, ``A simple improvement for permanent
  magnet systems for Kibble balances: More flat field at almost no cost,'' {\em
  IEEE Trans. Instrum. Meas.}, vol.69, no.~10, pp. 7752-7760, 2020.

\bibitem{schlamminger2013design}
S.~Schlamminger, ``Design of the permanent-magnet system for NIST-4,'' {\em
  IEEE Transactions on Instrumentation and Measurement}, vol.~62, no.~6,
  pp.~1524--1530, 2013.

\bibitem{LNEmag}
P.~Gournay, G.~Genev{\`e}s, F.~Alves, {\it et al},
  ``Magnetic circuit design for the {BNM} watt balance experiment,''
  {\em IEEE Trans. Instrum. Meas.}, vol.~54, no.~2, pp.~742--745, 2005.

\bibitem{li17}
S.~Li, F.~Bielsa, M.~Stock, A.~Kiss, and H.~Fang, ``Coil-current effect in
  Kibble balances: analysis, measurement, and optimization,'' {\em Metrologia},
  vol.~55, no.~1, pp.~75--83, 2017.

\bibitem{hysteresis}
S.~Li, F.~Bielsa, M.~Stock, A.~Kiss, and H.~Fang, ``An investigation of
  magnetic hysteresis error in Kibble balances,'' {\em IEEE Trans. Instrum.
  Meas.}, 2019.

\bibitem{shisong2022}
S.~Li and S.~Schlamminger, ``The irony of the magnet system for {K}ibble
  balances -- a review,'' {\em Metrologia}, vol.~59, p.~022001, 2022.

\bibitem{li18}
S.~Li, M.~Stock, F.~Biesla, A.~Kiss, and H.~Fang, ``Field analysis of a moving current-carrying coil in OMOP Kibble balances,'' in {\em 2018 International Applied Computational Electromagnetics Society Symposium (ACES)}, pp.~1--2,
  IEEE, 2018.

\bibitem{BIPMmag2017}
S.~Li, F.~Bielsa, M.~Stock, A.~Kiss, and H.~Fang, ``A permanent magnet system
  for Kibble balances,'' {\em Metrologia}, vol.~54, no.~5,
  pp.~775--783, 2017.

\bibitem{kibble2014principles}
B.~Kibble and I.~Robinson, ``Principles of a new generation of simplified and
  accurate watt balances,'' {\em Metrologia}, vol.~51, no.~2, p.~S132, 2014.

{
\bibitem{UME2}
H. Ahmedov, R. Orhan and B. Korutlu, ``UME Kibble balance operating in air,'' {\em Metrologia}, vol. 60, no. 1, p. 015003, 2023.

\bibitem{nimexmag}
J. Xu, Q. You, Z. Li, {\em et al},
``Research on the effect of the external 
magnetic field in the joule balance at NIM,''  {\em Metrologia}, vol. 55, no. 3, pp. 392-403, 2018.
}

\bibitem{BIPMmagShielding}
S.~Li, M.~Stock, F.~Bielsa, A.~Kiss, and H.~Fang, ``Shielding performance
  evaluation of BIPM-type Kibble balance magnet,'' in {\em 2018 Conference on
  Precision Electromagnetic Measurements (CPEM 2018)}, pp.~1--2, IEEE, 2018.

\end{thebibliography}

\end{document}